\documentclass[floats,floatfix,showpacs,amssymb,prd,twocolumn,superscriptaddress,nofootinbib]{revtex4-1}

\usepackage{amssymb,amsmath,verbatim,mathtools,needspace,enumitem,etoolbox,graphicx,physics,microtype,afterpage,bm}

\usepackage{subcaption}
\usepackage{ragged2e}
\DeclareCaptionJustification{justified}{\justifying}
\usepackage[dvipsnames, usenames]{xcolor}
\usepackage[normalem]{ulem}

\definecolor{linkcolor}{rgb}{0.0,0.3,0.5}
\usepackage[unicode, colorlinks=true, linkcolor=linkcolor, citecolor=linkcolor, filecolor=linkcolor,urlcolor=linkcolor, pdfusetitle]{hyperref}
\usepackage[all]{hypcap}
\usepackage[T1]{fontenc}
\usepackage[utf8]{inputenc}
\usepackage{tabularx}
\usepackage{float}
\allowdisplaybreaks
\graphicspath{{./figure/}}

\newcommand{\lp}{\left (}
\newcommand{\rp}{\right )}

\newcommand{\be}{\begin{equation}\begin{aligned}}
\newcommand{\ee}{\end{aligned}\end{equation}}

\newcommand{\bbe}{\begin{align}}
\newcommand{\eee}{\end{align}}

\newcommand{\bea}{\begin{eqnarray}}
\newcommand{\eea}{\end{eqnarray}}

\def\beq{\begin{equation}}
\def\eeq{\end{equation}}

\def\d{{\rm d}}

\def\beqa{\begin{eqnarray}}

	\def\eeqa{\end{eqnarray}}
	
\def\lsim{\mathrel{\rlap{\lower4pt\hbox{\hskip0.5pt$\sim$}}
		\raise1pt\hbox{$<$}}}     
\def\gsim{\mathrel{\rlap{\lower4pt\hbox{\hskip0.5pt$\sim$}}
		\raise1pt\hbox{$>$}}}     

\def\d{{\rm d}}

\def\d{{\rm d}}

\def\M{{\tiny M}}

\def\M{{{M}}}

\def\eeqa{\end{eqnarray}}

\def\bq{\begin{quote}}
\def\eq{\end{quote}}

 at 10truept

\def\eeqa{\end{eqnarray}}
\def\lsim{\mathrel{\rlap{\lower4pt\hbox{\hskip0.5pt$\sim$}}
  \raise1pt\hbox{$<$}}}     
\def\gsim{\mathrel{\rlap{\lower4pt\hbox{\hskip0.5pt$\sim$}}
  \raise1pt\hbox{$>$}}}     

\definecolor{rb4}{HTML}{27408B}

\begin{document}

\title{Bubble Correlation in First-Order Phase Transitions}

\author{Valerio~De~Luca}
\email{Valerio.DeLuca@unige.ch}
\affiliation{D\'epartement de Physique Th\'eorique and Centre for Astroparticle Physics (CAP), Universit\'e de Gen\`eve, 24 quai E. Ansermet, CH-1211 Geneva, Switzerland}

\author{Gabriele~Franciolini}
\email{Gabriele.Franciolini@unige.ch}
\affiliation{D\'epartement de Physique Th\'eorique and Centre for Astroparticle Physics (CAP), Universit\'e de Gen\`eve, 24 quai E. Ansermet, CH-1211 Geneva, Switzerland}

\author{Antonio~Riotto}
\email{Antonio.Riotto@unige.ch}
\affiliation{D\'epartement de Physique Th\'eorique and Centre for Astroparticle Physics (CAP), Universit\'e de Gen\`eve, 24 quai E. Ansermet, CH-1211 Geneva, Switzerland}

\date{\today}

\begin{abstract} \noindent
Making use of  both the stochastic approach to the tunneling phenomenon and the threshold statistics, we offer a simple argument to show that critical bubbles may be correlated in first-order phase transitions and  biased compared to the underlying scalar field spatial distribution. This happens though only if  the typical  energy scale of the phase transition is sufficiently high.  We briefly discuss possible implications of this result, e.g. the formation of primordial black holes through bubble collisions.
\end{abstract}

\maketitle

\paragraph*{\it 1. Introduction.} First-order phase transitions in the early universe \cite{Hindmarsh:2020hop} play a crucial role in cosmology as they may have left behind detectable relics such as gravitational waves \cite{Caprini:2018mtu}, the baryon asymmetry \cite{Riotto:1999yt} and the formation of primordial black holes \cite{Hawking:1982ga}. These phenomena depend crucially on the typical mean distance between the bubbles of the broken phase which end up  occupying all the available volume triggering the end of the transition. For instance, both the amplitude and the peak frequency of the gravitational waves generated
at phase transitions depend on such mean distance \cite{Caprini:2018mtu}.

Through $(1+1)$-dimensional real-time models at zero temperature \cite{Braden:2018tky},  Ref.~\cite{Pirvu:2021roq}  has recently  shown  that bubbles are not only clustered, but that the bubble sites  follow the statistics of maxima of a Gaussian field \cite{Bardeen:1985tr}. Inspired by such a result, in this short note we provide   a simple and intuitive argument 
-- based on the stochastic approach to tunneling and threshold statistics -- showing why bubbles may be indeed clustered and biased with respect to the underlying scalar field spatial distribution. However, we will argue that  bubble correlation is relevant only if the typical energy scale of the problem is very high (much larger than the electroweak scale). Clustering is indeed not an automatic property:   one needs to evaluate the clustering length and assure that the number of bubbles in a volume of size the clustering length is (much) larger than unity. This turns out to be the case only 
for first-order phase transitions happening either at very high values of the temperature or Hubble rate. 
 While our results do not have  implications for the predictions of the gravitational wave spectrum, they  may be relevant  for other considerations, e.g. for the generation of primordial black holes.

\paragraph*{\it 2. Critical bubble nucleation through the stochastic approach.}
\noindent
Our starting and benchmark model  is the  one-loop effective scalar field potential which can be written at finite temperature $T$ as \cite{Quiros:1999jp}
\be
V(\phi, T) = \frac{ m^2(T)}{2} \phi^2 - E T \phi^3 + \frac{\lambda}{4} \phi^4.
\ee
We will discuss a benchmark potential at zero temperature at the end of this section.

In the following, we will adopt the thermal mass parametrisation as
\begin{equation}
	m^2(T) = 2 D \lp T^2-T_0^2\rp,
\end{equation}
where the reference temperature $T_0$ is defined as the temperature above which the scalar field does not posses a metastable minimum at the origin and $D$ is a model dependent constant.

The most   suitable way  to
directly and intuitively  study the bubble correlation function is to adopt 
 the  stochastic approach to tunnelling (see for example \cite{Linde:1991sk,Linde:1990flp,Dine:1992wr, Hindmarsh:1993mb}). 
 The first step is to characterise the bubbles which are able to expand. 
In the limit of small bubble velocity and assuming O(3) spherical symmetrical bubble solutions, the equation of motion describing the evolution of
the field $\phi$ at finite temperature is
\begin{equation}\label{32}
\ddot\phi = \d^2\phi/\d r^2 + (2/r)\d\phi/\d r  - {\rm d} V/\d\phi.
\end{equation}
In order to have an energetically favourable process, the 
field $\phi$ inside a critical bubble should be larger than $\phi_*$, defined as $V(\phi_*,T) = V(0,T)$.
This also automatically implies that the field will be beyond the barrier 
where the effective potential has a negative derivative ${\rm d} V/d\phi < 0$.
Following Ref.~\cite{Dine:1992wr}, we neglect the quartic coupling and find $\phi_* = m^2/2ET$. To be conservative, we  will require 
field values beyond   $\phi_\text{\tiny th} \sim  2\phi_* = m^2/ET$.
The requirement of having expanding bubble $\ddot\phi > 0$ translates into
\begin{equation}\label{33}
 |\d^2\phi/\d r^2 + (2/r)\d\phi/\d r| < - V^\prime(\phi).
\end{equation}
This condition implies that the size of the bubble
should be sufficiently large not to have the gradient terms larger than the potential-induced drift $|V^\prime(\phi)|$.
Following again Ref. \cite{Dine:1992wr}, we can make a very rough estimate and write 
\begin{equation}\label{34}
{1\over 2} r^{-2} \lesssim  \phi^{-1}|V^\prime(\phi)| \sim 2 m^2 ,
\end{equation}
which in momentum space translates into the requirement $k \lesssim k_\text{\tiny max} \simeq  2 m(T)$.

Starting from the correlation in momentum space
\be
\langle \phi (\vec k_1) \phi (\vec k_2) \rangle = \frac{1}{\omega_{k_1}} \lp \frac{1}{2} + n_\text{\tiny B} (k_1) \rp (2 \pi)^3 \delta_D (\vec k_1 + \vec k_2),
\ee
where $\omega_{k_1} = \sqrt{k_1^2 + m^2}$, $\delta_D(\vec k)$ is the three-dimensional Dirac distribution and $n_\text{\tiny B}$ is the Bose-Einstein distribution, we can compute the dispersion of thermal fluctuations of the scalar field $\phi$, 
satisfying the expansion condition (i.e.  $k \lesssim k_\text{\tiny max}$), as 
\begin{align}
	\label{35}
\sigma^2\equiv \langle \phi^2 \rangle_{k<k_\text{\tiny max}} & \simeq
{1\over 2\pi^2}
\int_{0}^{k_\text{\tiny max}}
\hspace{-0.6 cm}
{k^2 \d k\over
\sqrt {k^2 + m^2} \left[\exp\lp{\sqrt{k^2 + m^2}\over T}\rp - 1
\right]} \nonumber \\
&\simeq{C^2   T m\over  \pi^2}.
\end{align} 
We have approximated the integral by evaluating the function close to the dominant domain $k\sim k_\text{\tiny max}$, which amounts to neglect the mass term in the expressions.  
Also, the constant $C$ is an $\mathcal{O}(1)$ coefficient introduced to 
track the uncertainties in the computation of $k_\text{\tiny max} $ and the integral \cite{Dine:1992wr}.
Assuming Gaussian statistics for the field thermal fluctuations, the field value distribution goes like
\begin{equation}\label{37} 
P(\phi)
\sim \exp\lp-{\phi^2\over 2\sigma^2 }\rp.
\end{equation}
Therefore, as the phase transition is generated by critical bubbles with field values larger than $\phi_\text{\tiny th}$, the relevant nucleation probability can be computed by integrating the distribution of field values above the threshold, that is
\begin{eqnarray}\label{Pabovet}
	P_1(\phi > \phi_\text{\tiny th} ) 
	&=& \int_{\phi_\text{\tiny th}}\d \phi\,  P(\phi)
	\sim \exp\lp-\phi_\text{\tiny th}^2/ 2\sigma^2 \rp \nonumber\\ 
	&\sim& 
	\exp \lp-{
m^3\pi^2\over  2 C^2 E^2 T^3}\rp,
\end{eqnarray}
where we adopted the large threshold limit and neglected the prefactor. 
Quite remarkably, this simple  approach recovers the result one obtains by computing the tunnelling probability by evaluating the Euclidean bounce solution (where $S_3$ is the three-dimensional Euclidean action) \cite{Linde:1981zj} 
\begin{equation}\label{19}
P_1 \simeq \exp\lp - {S_3\over T}\rp \sim \exp \lp-{4.85
m^3\over E^2 T^3}\rp,
\end{equation}
by setting $C^2 = 1.02$, value which we assume from now on.

Provided that $\lambda D-E^2>0$, a second energetically favoured minimum in the potential appears when the temperature drops below a critical value. This can be found by evaluating the temperature allowing the second minimum to be degenerate with the false vacuum, which is~\cite{Dine:1992wr}
\be
T_c^2 = \frac{T_0^2}{1-E^2/\lambda D},
\ee
such that the corresponding thermal mass is given by
\be
\frac{m^2 (T_c)}{T_c^2} = \frac{2 E^2}{\lambda}.
\ee
Finally, we notice that the strength of the phase transition can be parametrised in terms of the order parameter $\phi (T_c)/T_c = 2 E/ \lambda$, which implies that strong phase transitions are obtained for small values of $\lambda$. Furthermore, one can determine the inverse time duration $\beta$ and the ratio of the vacuum to the radiation energy density $\alpha$ as~\cite{Enqvist:1991xw}
\begin{align}
\frac{\beta}{H(T_c)} &\equiv T \frac{\d}{\d T}\lp \frac{S_3}{T} \rp \bigg|_{T = T_c} =3 \sqrt{2}\pi^2 \frac{(\lambda D - E^2)}{E \lambda^{3/2}},
\nonumber \\
\alpha &\equiv 
\left .\frac{\rho_\text{\tiny vac}}{\rho_\text{\tiny rad}}\right |_{T = T_c}  = \frac{60}{\pi^2 g_*} \frac{E^2( \lambda D-E^2)}{\lambda},
\end{align}
in terms of the Hubble parameter at the transition temperature 
$H(T_c)$,  the vacuum energy density associated with the transition
$\rho_\text{\tiny vac}(T_c) =(\Delta V -T \d V/\d T)|_{T=T_c}$, the radiation energy density $\rho_\text{\tiny rad} (T_c) = \pi^2 g_* T_c^4/30$ and the effective number of degrees of freedom $g_*$. Notice that the first term in $\rho_\text{\tiny vac}(T_c)$ vanishes when computed at $T=T_c$ since the stable and metastable minima are degenerate at the critical temperature.
Small values of $\lambda$ (consistently with the previous assumption) implies large ratio between the vacuum and radiation energy density and a shorter duration of the phase transition.

As promised, let us consider as well a benchmark potential at zero temperature
\be
V(\phi)=\frac{M^2}{2}\phi^2-\frac{\delta}{3}\phi^3+\frac{\lambda}{4}\phi^4.
\ee
We adopt the stochastic tunneling approach to deal with the fluctuations of  quantum nature. Neglecting again for simplicity the quartic coupling, one can estimate $\phi_\text{\tiny th}\simeq 3M^2/\delta$   and 
\begin{align}
	\label{zero}
\sigma^2& =
{1\over 2\pi^2}
\int_{0}^{k_\text{\tiny max}}
{k^2 \d k\over
\sqrt {k^2 + M^2} } \nonumber\\
&=\frac{M^2}{4 \pi^2}\left[C\sqrt{1+C^2}
-{\rm arc}\,{\rm  sinh} \, C \right],
\end{align} 
where we have parametrized
$k_\text{\tiny max}=CM$. One can easily check that the corresponding tunneling rate
computed using the threshold statistics matches the known result, in terms of the four-dimensional Euclidean action $S_4$, $P_1\simeq {\rm exp}\left(-S_4\right) \sim  {\rm exp}\left(-205 \M^2/\delta^2\right)$\cite{Linde:1990flp} for $C\simeq 1.2$, again a remarkable result. The approximate duration of the phase transition $1/\beta$ can be estimated by imposing the fraction of volume in the true vacuum at the end of the transition to be order unity, that is $\Gamma/\beta^4 \sim 1$~\cite{Kosowsky:1992rz,Turner:1992tz}. Given that the tunneling rate density is
$\Gamma \sim M^4\cdot P_1$, one finds $\beta/H \sim (M/H) \, {\rm exp}\left(-S_4/4 \right)\sim (M_\text{\tiny pl}/M)\left(-S_4/4 \right)$.

\paragraph*{\it 3. Bubble correlation function.}
Inspired by the simple description of  the previous section, in order to describe the bubble two-point correlation function $\xi_\text{\tiny b}(r)$ as a function of the distance $r$, we  adopt the threshold statistics routinely used in   large-scale structure. We will make use of threshold statistics instead of peak theory based on the standard argument in cosmology and galaxy bias that the two statistics deliver the same results in the limit of large thresholds.

If normalised with respect to the average number density, 
the spatial number density  of discrete bubble nucleation centers at position $\vec r_i$ is 
\begin{eqnarray}
\delta_\text{\tiny b}(\vec r)=\frac{1}{\bar n_\text{\tiny b}}\sum_i \delta_D(\vec r-\vec r_i)-1,
\end{eqnarray}
where $\bar n_\text{\tiny b}$ is the  mean bubble number density and $i$ indicates the various initial positions of bubbles.
Analogously to the large-scale structure theory (see for example Ref.~\cite{Baldauf:2013hka}), the corresponding two-point correlation function must take the general form \cite{Ali-Haimoud:2018dau}
\begin{align}
\big\langle\delta_\text{\tiny b}(\vec r)\delta_\text{\tiny b}(\vec 0) \big\rangle 
=
\frac{1}{\bar n_\text{\tiny b}}\delta_D(\vec r)+ \xi_\text{\tiny b}(r).
\label{eq:PBH2pt}
\end{align}
The first piece is due to the Poisson shot noise while $\xi_\text{\tiny b}(r)$ is the   (reduced) bubble  correlation function. 
At small scales,  roughly identified with the critical bubble size $r_\text{\tiny cr}$, the correlation function must respect the exclusion requirement arising because distinct bubbles cannot form arbitrarily close to each other.
As a result, the conditional probability to find a bubble  at a distance $r$ from another one, which is proportional to $1+\xi_\text{\tiny b}(r)$, must vanish for $r\lesssim r_\text{\tiny cr}$. 
Therefore, the reduced correlation becomes 
\be
\xi_\text{\tiny b}(r)\approx -1 \quad \mbox{for}\quad r\lesssim r_\text{\tiny cr},
\ee
which implies that bubbles are anti-correlated at short distances.

We now compute the correlation function of critical bubbles. Using the stochastic approach adopted in the preceding steps, one can compute the correlation of nucleation sites by computing the correlation function of 
the scalar field values above the threshold $\phi_\text{\tiny th}$. In other words, we will be using the same threshold statistics cosmologists  are familiar with when studying the galaxy correlators: as galaxies (or, better to say, dark matter halos where galaxies end up) are discrete objects formed when the dark matter overdensity is above a given threshold value and the galaxy correlators are biased with respect to the dark matter ones \cite{Desjacques:2016bnm}, in the very same way critical bubbles are formed only when the scalar field is above a given threshold, and therefore critical bubbles will be biased compared to the underlying scalar field spatial distribution.

In general, the threshold correlation function is defined as 
\begin{equation}
	1 + \xi_\text{\tiny b}(r) = \frac{P_2}{P_1^2}, 
\end{equation}
where $P_1$ is probability of one region being above threshold, given by Eq.~\eqref{Pabovet}, 
and $P_2$ is the probability that two regions separated by $r$ are both above threshold \cite{Kaiser:1984sw} 
\begin{equation}
	P_2 =  \int_{\nu}^\infty \frac{\d x_1 \d x_2}{2 \pi}   \frac{1}{\sqrt{1- w^2}} 
\exp\left[ - \frac{x_1^2 + x_2^2 - 2 w x_1 x_2}{2(1 - w^2)} \right], 
\end{equation}
in terms of the adimensional threshold $\nu \equiv \phi_\text{\tiny th} / \sigma$, field values $x_i \equiv \phi(r_i) / \sigma$ and 
rescaled field correlation function
$w (r) = \xi_\phi (r)/ \sigma^2$.
The threshold correlation function takes the general form 
\begin{equation}
1 + \xi_\text{\tiny b}(r) \approx (1 + w)   \frac{\textrm{erfc}\left( \sqrt{\frac{1 - w}{1 + w}}~ \nu/\sqrt{2}\right)}{\textrm{erfc}(\nu/\sqrt{2})}
\quad \text{for} \quad  \nu \gg 1,
\end{equation}
while it reduces to the well-known result  \cite{Kaiser:1984sw} 
\begin{equation}
	\xi_\text{\tiny b} (r)
	\equiv  b_1^2 \xi_{\phi} (r)
	\simeq 
	\lp \frac{\phi_\text{\tiny th}}{\sigma^2 } \rp^2 \xi_{\phi} (r),
\end{equation}
in the limit of $w \ll 1/\nu^2 \ll 1$.
The physical intuition for strong first-order phase transition leading to clustered bubbles can be obtained by noticing that the bias factor is proportional to the order parameter, such that its larger values result into a stronger bubble correlation.
The scale independent bias factor in the previous expression takes the form at finite temperature
\begin{equation}
	b_1^2 = \frac{\pi^4 m^2}{ E^2 T^4},
\end{equation}
which at the temperature $T_c$ takes the value
\begin{equation}
	b_1^2(T_c) \sim \frac{2\pi^4}{  \lambda T_c^2}\gg \frac{1}{T_c^2}.
\end{equation}
Finally,  the scalar field correlation function is
\begin{align}
\xi_\phi (r) & = 
{1\over 2\pi^2}\int_{0}^{k_\text{\tiny max}}  \d k k^2
{ j_0 (k r)  \over
\sqrt {k^2 + m^2} \left[\exp\lp{\sqrt{k^2 + m^2}\over T}\rp - 1
\right]}
.
\end{align} 
The spherical Bessel function is constant $j_0 (k r) \sim 1$ up to momentum values $k r < 1$, and decreases rapidly otherwise. 
In the relevant limit when $r \gsim 1/k_\text{\tiny max} \simeq 1/2m$, one can neglect the momentum scale with respect to the scalar field thermal mass and find
\begin{align}
	\xi_\phi (r) & \simeq
	{1\over 2\pi^2}\int_{0}^{k_\text{\tiny max}}
	{k^2 \d k\over m}\left[\exp\lp{m\over T}\rp - 1
	\right]^{-1} j_0(k r ) 
	\nonumber \\
	& \simeq
	\frac{2T m}{\pi^2}
	\frac{j_1(2 m r)}{ m r }.
\end{align}
We do not consider the opposite regime $r<1/k_\text{\tiny max}$ as it would correspond to distances smaller than the minimum size of an expanding bubble. 

One can repeat the same exercise at zero temperature with the potential (\ref{zero}) and find
\begin{eqnarray}
b_1^2 &\simeq&\frac{192 \pi^4}{\delta^2},\nonumber\\
	\xi_\phi (r) &\simeq &
	\frac{M^2}{2\pi^2}
	\frac{j_1( M r)}{ M r }.
\end{eqnarray}
Notice that small values of $\delta$ lead to a strong phase transitions and large bias.

\paragraph*{\it 4. The bubble clustering scale.}
One can estimate the clustering length imposing  $	\xi_\text{\tiny b} (r) \sim 1$. 
This requirement can be understood by considering the counts of neighbours. 
The mean count of bubble nucleation sites $\langle N\rangle$ in a cell of volume $V$ centered on a bubble is
\be
\label{eq:meancount}
\langle N\rangle = \bar n_\text{\tiny b} V + \bar n_\text{\tiny b} \int_V\!{\rm d}^3r\,\xi_\text{\tiny b}(r).
\ee
The mean count $\langle N\rangle$ significantly deviates from Poisson if the contribution from the second piece in Eq.~\eqref{eq:meancount}
rises above the discreteness noise $\bar n_\text{\tiny b} V$.
This can happen at scales $r \lesssim r_\text{\tiny cl}$ defined as the characteristic  clustering length through the relation $\xi_\text{\tiny b}(r_\text{\tiny cl})=1$.

Using the large scale limit for the peak correlation function and neglecting the periodic oscillations induced by the Bessel function,  one has
\begin{equation}
	\lp \frac{\phi_\text{\tiny th}^2}{\sigma^2 } \rp  (r_\text{\tiny cl}  k_\text{\tiny max})^{-2}  \sim 1,
\end{equation}
which gives
\be
r_\text{\tiny cl} =\frac{\pi}{2^{1/4}} \lp \frac{E^{1/2}}{\lambda^{3/4}} \rp \frac{1}{m},
\ee
at finite temperature and
\be
r_\text{\tiny cl} =\frac{6 \sqrt{2} \pi}{3^{1/4}\delta},
\ee
at zero temperature.

\begin{figure}[t]
	\centering
	\includegraphics[width=0.9 \linewidth]{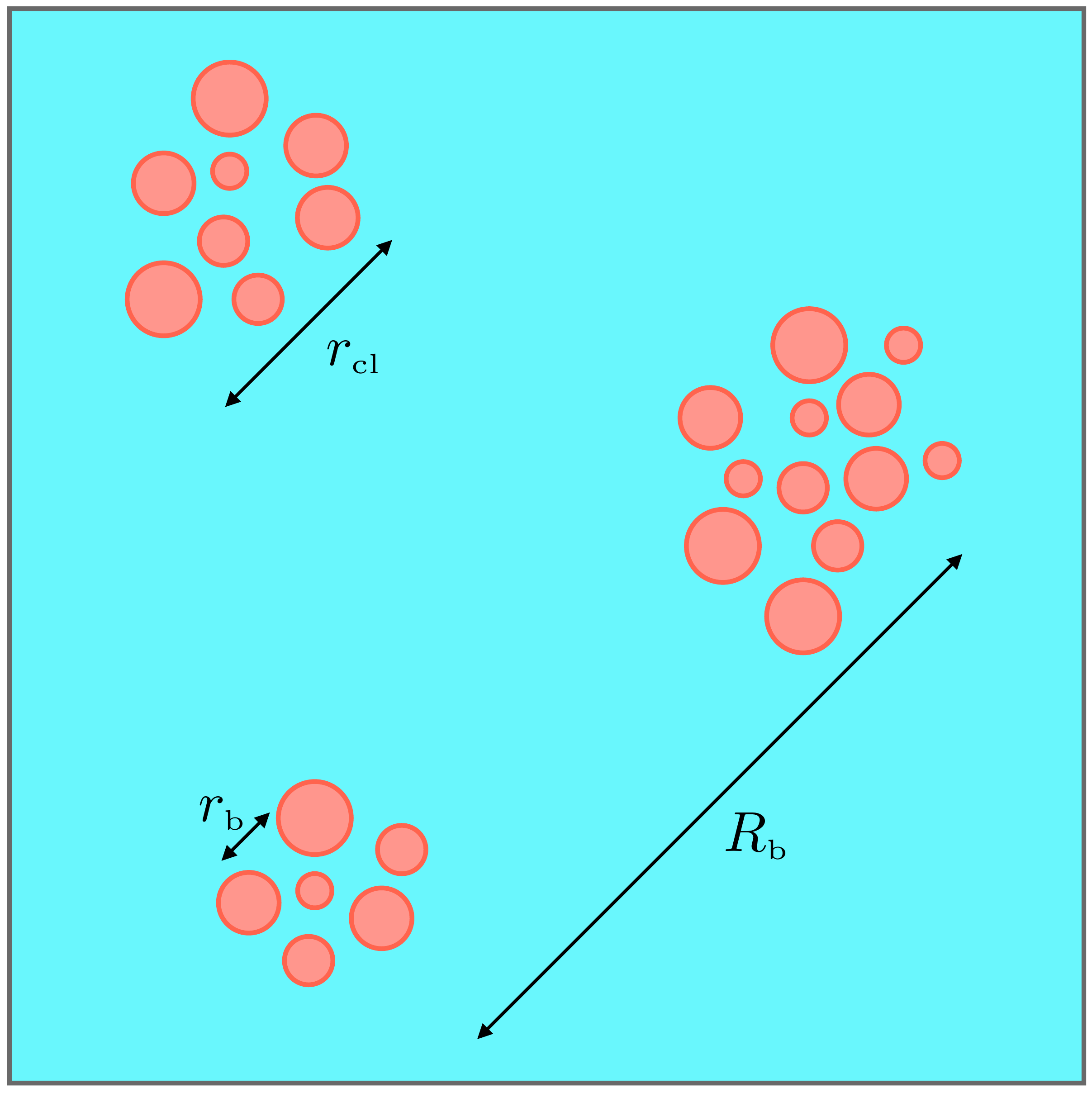}
	\caption{\it Pictorial representation of bubble clustering with the different characteristic length scales of the problem.
	}
	\label{fig:1}
\end{figure}

\paragraph*{\it 5. When are bubbles correlated?} 
\noindent
One should not claim victory too soon, though. The existence of a clustering length does not imply automatically that correlation is relevant. Indeed, let us inspect Eq.~\eqref{eq:meancount}. The Poisson contribution increases like the volume centered around a given bubble and therefore scale like $\sim r^3$. On the other hand, the piece from the correlation function scales like $\sim r$, as the correlation function scales like $\sim 1/r^2$. While the two contributions become of the same order at $\sim r_\text{\tiny cl}$, one has to also estimate the average number of bubbles in a volume of size  the correlation length, that is
\be
\langle N\rangle \sim \bar n_\text{\tiny b}r_\text{\tiny cl}^3 \sim \frac{1}{v^3} \left(\beta r_\text{\tiny cl}\right)^3= \frac{1}{v^3} \lp \frac{\beta}{H} \rp^3 \lp \frac{r_\text{\tiny cl}}{H^{-1}} \rp^3, 
\ee
where $v$ is the bubble velocity. 
At finite temperature, the mean number reduces to ($M_\text{\tiny pl}$ being the Planck mass)
\be
\langle N\rangle \sim \frac{1}{v^3} \frac{(\lambda D - E^2)^{3}}{E^{9/2}\lambda^{21/4}} \lp \frac{T_c}{M_\text{\tiny pl}} \rp^3,
\ee
while at zero temperature it becomes
\be
\langle N\rangle \sim \frac{1}{v^3} 
\lp \frac{M}{\delta} \rp^6
\lp \frac{H}{\delta} \rp^3
.
\ee
From these expressions one immediately recognizes that the average number of bubbles is larger than unity at the cluster scale only if  we are dealing with high-energy phase transitions. In the finite temperature case, for $v\sim 0.1$, $D\sim E\sim \lambda\sim 10^{-2}$, corresponding to $(\beta/H)\sim 10^2$, we get 
\be
\label{aa}
T_c\gtrsim 10^{-5}\left(\frac{10^2}{\beta/H}\right)M_\text{\tiny pl}\simeq \left(\frac{10^2}{\beta/H}\right) 10^{13}\, {\rm GeV},
\ee
while for the zero temperature case one finds, for $v\sim 1$ and $\delta\sim \epsilon  M$ (with $\epsilon$ a coefficient smaller than unity),  $H\sim M^2/M_\text{\tiny pl} \gtrsim \epsilon^3 M$, or 
\be
\label{con}
M\gtrsim \epsilon^3M_\text{\tiny pl}.
\ee
These tight requirements for bubble clustering are  essentially due to the fact that the average number density of bubbles tends to be  
small in a volume with size the cluster length, the typical bubble distance being still a sizeable fraction of   the Hubble radius and therefore much larger than the cluster length. 
A useful analogy may be drawn with the clustering of primordial black holes at formation: being the appearance of primordial black holes a rare event, i.e. their average number density is very small, their average number in a volume of size the cluster length is smaller than unity. Primordial black holes are therefore not clustered at formation being rare phenomena \cite{Desjacques:2018wuu}
and 
critical expanding bubbles are subject to the same reasoning. Notice also that the effect of bubble clustering is more  relevant for strong phase-transitions whose duration is short, in agreement with the findings of Ref.~\cite{Pirvu:2021roq}.

\paragraph*{\it 6. Possible implications and conclusions.} 
\noindent
One can envisage several applications  of our findings. The most immediate one regards the production of gravitational waves during phase transitions, see for instance  Refs. \cite{Caprini:2018mtu,Pirvu:2021roq}. Gravitational waves may be produced by several mechanisms like bubble collisions, turbulence and sound waves in the vicinity of the bubble walls. 
All of them are sensitive to the mean bubble distance both in the amplitude and in the peak frequency.  If bubbles are strongly correlated at nucleation in peculiar sites, they will collide with a typical mean distance $r_\text{\tiny b}$ which is smaller than the Poisson case $R_\text{\tiny b}$. 
From Eq.~\eqref{eq:meancount} we find 
\begin{equation}
r_\text{\tiny b}\sim \frac{1}{(\bar n_\text{\tiny b}\xi_\text{\tiny b})^{1/3}}\ll \frac{1}{\bar n_\text{\tiny b}^{1/3}} \sim R_\text{\tiny b}.
\end{equation}
Subsequently the few bubbles resulting from the collision of the clustered ones will collide at a larger distance $R_\text{\tiny b}\sim \bar n_\text{\tiny b}^{-1/3}$, see Fig.~\ref{fig:1} for a pictorial representation.
Even though these considerations are probably oversimplified, we expect that the gravitational  wave spectrum has two peaks at $f_\text{\tiny GW}\sim 1/r_\text{\tiny b} $ and $\sim 1/R_\text{\tiny b}$, separated by an amount $\sim \xi_\text{\tiny b}^{1/3}$, with the amplitude of the gravitational wave at $f_\text{\tiny GW}\sim 1/r_\text{\tiny b}$ smaller than the one at $f_\text{\tiny GW}\sim  1/R_\text{\tiny b}$, being the amplitude directly proportional to powers of the bubble mean distance for all mechanisms.   However, plugging numbers in, our results indicate that for clustering to be important one would need to go to very high gravitational wave frequencies, i.e. in the GHz range where detecting gravitational waves is rather challenging \cite{Aggarwal:2020olq}\footnote{ The typical frequency of the gravitational wave scales proportionally to $(\beta/H) T_c$ \cite{Caprini:2018mtu} and therefore does not change modifying $(\beta/H)$  when the condition (\ref{aa}) is imposed.}.

On a more positive side, bubble clustering may help the formation of primordial black holes
by bubble collisions \cite{Hawking:1982ga, Khlopov:1998nm, Jung:2021mku} in supercooled phase transitions. Primordial black holes may form if a large enough energy is deposited
in bubble walls and such an energy is concentrated within its corresponding Schwarzschild radius. This requires the collision of a large enough number of bubbles for which clustering may help.  Following Ref.~\cite{Hawking:1982ga}, the condition to form a primordial black hole in a volume of size the clustering length is 
\be
\langle N\rangle \gtrsim \frac{4}{r_\text{\tiny cl} H},
\ee
which implies
\be
M\lesssim \frac{H}{\epsilon^{5/2}}\,\,\,\,{\rm or}\,\, \,\,
M\gtrsim \epsilon^{5/2} M_\text{\tiny pl},
\ee
which, as expected, is a condition parametrically more stringent than the one in Eq.~\eqref{con}. Such primordial black holes would have however a typical mass of the order of $\sim M_\text{\tiny pl}/\epsilon^6\langle N\rangle$ and therefore quickly evaporate.

A large bubble correlation may also induce a change in the dynamics of the bubble collisions in the scenario of eternal inflation, potentially affecting their
observational signatures in the CMB~\cite{Aguirre:2009ug}.

Finally, we  mention the possibility of inducing the phase transition by accumulating bubbles with radii below the critical values. We expect such an induction to be favoured by a strong bubble clustering.

 \vspace{.5 cm}
\paragraph*{ Acknowledgments.}
We are indebted with J. Braden, M. Hindmarsh, M. Johnson and D. Pirvu for comments on the draft. We also thank C. Caprini for discussions on the GW signals produced by first-order phase transitions and  V. Desjacques for  discussions on the clustering.
V.DL., G.F. and A.R. are supported by the Swiss National Science Foundation 
(SNSF), project {\sl The Non-Gaussian Universe and Cosmological Symmetries}, project number: 200020-178787.

\bibliography{main}

\end{document}